\documentclass[iop]{emulateapj}

\newcommand{\NH}{N_\mathrm{H}}
\newcommand{\cost}{\cos\theta}

\usepackage{graphicx}
\usepackage{CJK}
\shorttitle{An X-ray model for clumpy tori} \shortauthors{Liu \& Li}

\begin{document}
\begin{CJK*}{GB}{gbsn}

\title{An X-ray spectral model for clumpy tori in active galactic nuclei}


\author{Yuan Liu (ÁõÔª) and Xiaobo Li (ÀîС²¨)}

\affil{Key Laboratory of Particle Astrophysics, Institute of High
Energy Physics, Chinese Academy of Sciences, P.O.Box 918-3, Beijing
100049, China}

\email{liuyuan@ihep.ac.cn; lixb@ihep.ac.cn}

\begin{abstract}
We construct an X-ray spectral model for the clumpy torus in an
active galactic nucleus (AGN) using Geant4, which includes the
physical processes of the photoelectric effect, Compton scattering,
Rayleigh scattering, $\gamma$ conversion, fluorescence line, and
Auger process. Since the electrons in the torus are expected to be
bounded instead of free, the deviation of the scattering cross
section from the Klein-Nishina cross section has also been included,
which changes the X-ray spectra by up to 25\% below $10$ keV. We
have investigated the effect of the clumpiness parameters on the
reflection spectra and the strength of the fluorescent line Fe
K$\alpha$. The volume filling factor of the clouds in the clumpy
torus only slightly influences the reflection spectra, however, the
total column density and the number of clouds along the line of
sight significantly change the shapes and amplitudes of the
reflection spectra. The effect of column density is similar to the
case of a smooth torus, while a small number of clouds along the
line of sight will smooth out the anisotropy of the reflection
spectra and the fluorescent line Fe K$\alpha$. The smoothing effect
is mild in the low column density case ($\NH=10^{23}$ cm$^{-2}$),
whereas it is much more evident in the high column density case
($\NH=10^{25}$ cm$^{-2}$). Our model provides a quantitative tool
for the spectral analysis of the clumpy torus. We suggest that the
joint fits of the broad band spectral energy distributions of AGNs
(from X-ray to infrared) should better constrain the structure of
the torus.
\end{abstract}



\section{Introduction}
In the unified model of active galactic nuclei (AGNs), a dusty torus
is proposed to account for the apparent difference between type 1
and 2 AGNs, i.e., the central source is obscured by the torus in
type 2 AGNs but not in type 1 AGNs (Antonucci 1993; Urry \& Padovani
1995). Despite the key ingredient of the unification model, the
origin and structure of the torus are still not clear. An important
debate about the torus structure is whether the material in a torus
is smooth or clumpy (Feltre et al. 2012; H\"{o}nig 2013). Both
models have their advantages and drawbacks. The torus absorbs
high-energy photons from an accretion disk/corona and converts them
into the infrared band. Therefore, it is possible to constrain the
structure of a torus by its infrared spectrum. The smooth model
predicts strong silicate features in the infrared band; however, the
observed feature is much weaker than the model's prediction (Laor \&
Draine 1993; Nenkova et al. 2002; Nikutta et al. 2009). The clumpy
model can naturally explain the weakness of the silicate feature but
cannot produce sufficient near-infrared emission from hot dust (Mor
et al. 2009; Vignali et al. 2011). For a few nearby sources, the
near-infrared interferometers can directly constrain the structure
of the tori and indeed favor  the clumpy model (Tristram et al.
2007). However, this method is limited by the low sensitivity of the
infrared interferometer and cannot be applied to a large sample. By
fitting the infrared energy spectral distribution with a specific
model (smooth or clumpy), we can also obtain the parameters of the
particular model. It has recently been claimed that the covering
factors of type 2 AGNs are systematically larger than those of type
1 AGNs (Elitzur 2012).

Besides the infrared emission, the torus will also absorb and
reflect the X-ray photons. Therefore, the X-ray spectrum can provide
independent constraints on the structure of a torus. There are many
works on the reflection spectra of different geometries. The
commonly used model \texttt{pexrav} in \textit{XSPEC} accounts for
the reflection from a flat disk with infinite optical depth
(Magdziarz
 \& Zdziarski 1995). However, fluorescence lines are not
self-consistently included in this model. The Gaussian lines should
be manually added in the fit model of real X-ray spectra. This work
is then improved by including the fluorescence line (e.g.,
\texttt{pexmon}), the ionization state of the reflection disk, and
 relativistic effects around a black hole (Nandra et al.
2007; Ross \& Fabian 2005; Brenneman \& Reynolds 2006). For the
smooth torus, detailed X-ray models are already available (Ikeda et
al. 2009; Murphy \& Yaqoob 2009; Brightman \& Nandra 2011),  though
the results are slightly different due to the discrepancies in
geometry, cross section, and simulation methods used. Some works
have also discussed the X-ray spectrum for clumpy media but do not
dedicate to tori (Nandra \& George 1994; Tatum et al. 2013). Yaqoob
(2012) discussed the qualitative effects on the reflection spectrum
and Fe K$\alpha$ line due to the clumpy structure.

It is necessary to utilize Monte Carlo simulations to deal with
clumpy geometry. In this paper, we present detailed results of
simulations on a clumpy torus. In Section 2, we will explain the
simulation process and other assumptions about the clumpy torus.
Then, the results of the reflection continuum and Fe K$\alpha$ are
presented in Sections 3 and 4, respectively. In Section 5, we
summarize our results and discuss the implications for observations.
As the first part of a series of papers, we will only discuss the
spectrum in this paper. The results of temporal response and
polarization will be given in subsequent papers.

\section{Simulation Method and assumptions}
We utilized an object-oriented toolkit, Geant4 (version
4.9.4),\footnote[1]{http://geant4.cern.ch/} to perform the
simulations. Three classes are necessary to construct the simulation
model in Geant4.

1 The geometry and material of the interaction region are described
by the class derived from G4VUserDetectorConstruction.

2 The class derived from G4VUserPhysicsList is required to construct
the particles and physical processes to be activated in the
simulation.

3 The class derived from G4VUserPrimaryGeneratorAction will generate
the primary events, including the type, energy, direction, and
position of the initial particles.

After the three classes have been defined, Geant4 will treat the
particles one by one and track the trajectories of primary particles
and secondary particles step by step. These particles will
participate in the physical processes activated in the simulation.
To determine the position of the interaction point of a given
physical process, Geant4 first calculates the mean free path
$\lambda$ as the function of energy
\begin{equation}\label{cs}
  \lambda(E)=(\sum_i[n_i\cdot\sigma(Z_i,E)])^{-1},
\end{equation}
where $n_i$ is the number density of the $i$th element,
$\sigma(Z_i,E)$ is the cross section per atom of the process, and
$\sum_i[]$ means the sum of all elements of the torus. Then the
number of the mean free paths the particle travels before the
interaction point is randomly determined by $-\log(\eta)$, where
$\eta$ is uniformly distributed in the range (0, 1). If the particle
is absorbed or escapes from the world boundary, the tracking of it
will be ended.\footnote[2]{The world boundary defines the largest
volume in Geant4 to contain all material. We define the world
boundary in our simulations as a box centered at the origin
\textit{O}  with side length  $10^{20}$ cm.} The result of the
relaxation of an exited atom is randomly determined by the atomic
data adopted, e.g., fluorescent yields. The secondary particles
(e.g., the Fe K$\alpha$ photons) are also tracked according to the
method above (Agostinelli et al. 2003).

At each step,  Geant4 will record the information of particles,
e.g., kinetic energy, momentum, position, time, and physical process
involved. Then the recorded information of every step can be used to
select the particles of interest. In the simulations in this paper,
photons that escape from the world boundary are selected to
construct the X-ray spectra in different situations. In the
following sections, we will discuss the assumptions in the three
classes of our simulations.

\subsection{Geometry and Constituents}
\begin{figure*}[htbp]
\plottwo{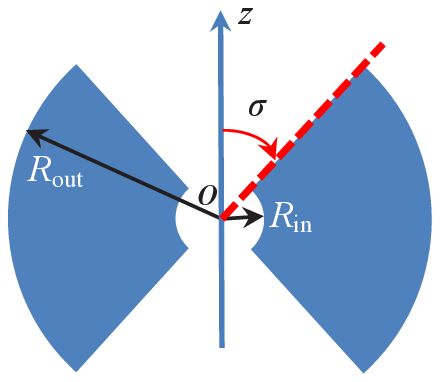}{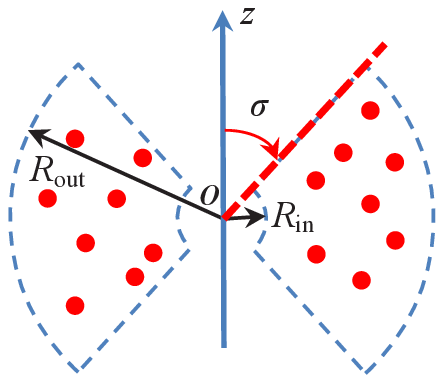}
 \caption{Cross-section view of the geometry of the smooth torus (left)
 and the clumpy torus (right) adopted in our simulations. For the
 smooth torus, the boundary is defined by the inner radius $R_\mathrm{in}$, the
 outer radius $R_\mathrm{out}$, and the half-opening angle $\sigma$; for the clumpy
 torus,
 the envelope of the clouds (red circles) is defined by the same parameters as for the smooth case.
 The central X-ray source is located at $O$. \label{fig1}}
\end{figure*}
For comparison, we also performed simulations of a smooth torus. The
geometry of the smooth torus is shown in Figure \ref{fig1} (left).
The boundaries are defined by the inner radius $R_\mathrm{in}=0.1$
pc, the outer radius $R_\mathrm{out}=2.0$ pc, and the half-opening
angle $\sigma=60^\circ$. We assume that the gas is uniformly
distributed in the torus. For the clumpy torus, we use the same
parameters ($R_\mathrm{in}$, $R_\mathrm{out}$, and $\sigma$) to
define the envelope of the torus, within which numerous spherical
clouds are uniformly distributed. In addition, we need parameters to
describe the clumpiness of the torus. There can be different choices
of the free parameters. In our simulation, we use the volume filling
factor $\phi$, the number of clouds along the line of sight $N$, and
total column density $\NH$ as the input parameters. Other
quantities, e.g., the radius of the cloud and the density of the gas
in the cloud, can be derived from these input parameters. $\phi$ and
$N$ determine the total number and the size of clouds and $\NH$
further determines the density of the gas in clouds. Figure
\ref{fig1} (right) is the configuration of the clumpy torus. In the
following simulations, we fix the parameters of the envelope, since
we focus on the comparison between the smooth and clumpy cases. It
is easy to change the parameters of the envelope in the subsequent
simulations if necessary. We further assume that the sizes of the
clouds are the same and the density of the gas in the cloud is
uniform. These assumptions are surely simplified compared with the
realistic tori of AGNs. However, the assumption ``uniform
distribution'' is widely used in the previous simulations about the
smooth torus. Hence, we follow this assumption and  mainly focus on
how the clumpiness influences the spectra. More complex and
realistic distributions will be investigated in future simulations.
We have included elements with the abundances from Anders \&
Grevesse (1989). The gas in the cloud is assumed to be cold, i.e.,
all atoms are in their ground states.

\subsection{Physical Processes}

\begin{figure}[htbp]
\plotone{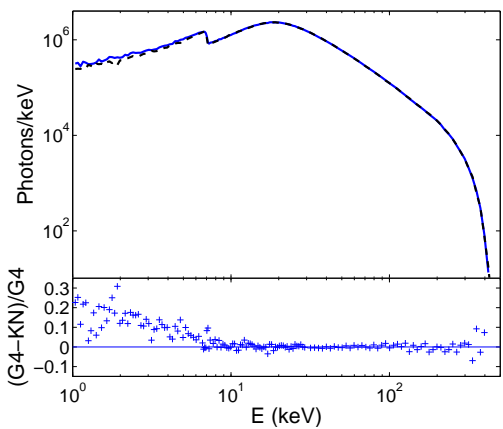}
 \caption{Comparison between spectra using the Klein-Nishina
cross section (black dashed line) and the corrected cross section
(blue solid line). The spectra are calculated with $\NH=10^{25}$
cm$^{-2}$, $N=2$, and $\phi=0.01$ in the face-on direction (see the
definition in Section 3). \label{figcom}}
\end{figure}

We invoked  the low-energy electromagnetic process in Geant4, which
is valid in the energy range from 0.25 keV to 100 GeV. For photons,
we considered the photoelectric effect, Compton scattering, Rayleigh
scattering, and $\gamma$ conversion. Fluorescence and Auger
processes were also loaded in the photoelectric effect. For
electrons, ionization, bremsstrahlung, and multiple scattering were
added in the process. The relevant cross sections and atomic data
are adopted from
EPDL97.\footnote[3]{https://www-nds.iaea.org/epdl97/} In spite of
slightly different cross sections of the photoelectric effect used
in previous simulations of torus, the most important difference is
the cross section of scatterings in our simulations. Since the
electrons in a torus are bounded, the Klein-Nishina cross section is
not appropriate (Hubbell et al. 1975). For the scattering by bounded
electrons, the cross section is divided into two parts, Rayleigh
scattering (coherent scattering, i.e., the atom is still in the
ground state after the scattering) and Compton scattering
(incoherent scattering, i.e., the atom is excited after the
scattering). The incoherent scattering dominates high-energy band
and tends to the Klein-Nishina cross section as the energy
increasing; while the coherent scattering is more important at
low-energy band and its cross section is proportional to the square
of the total number of the electrons in one atom.\footnote[4]{The
relative importance of the coherent and incoherent scattering is
graphically shown at
http://shape.ing.unibo.it/html/graphics\_of\_form\_factors.HTM.} As
a result, both the total cross section and the angular distribution
of scattered photons will deviate from the Klein-Nishina cross
section. The importance of this correction further depends on the
abundance of the gas. To evaluate the amplitude of the deviation in
the spectra, we replaced the cross section of scattering in Geant4
with the Klein-Nishina cross section and then compared the spectra
with that using the default cross section in Geant4. An illustration
of this comparison is shown in Figure \ref{figcom}. This correction
is indeed more important for low-energy photons. The deviation is
 negligible  above 10 keV and increases below 10 keV to 25\% at 1 keV. Under the current abundance, helium is responsible for most of the deviation in the spectra. Therefore,
this effect should be included in future simulations on the X-ray
spectra of tori.

\subsection{Incident Spectrum}
We adopted a single power law as the incident spectrum, i.e.,
$\mathrm{flux} \propto E^{-\Gamma}$ (1 keV $\leq E\leq$ 500 keV),
where $\Gamma$ is the photon index and fixed at 1.8 throughout the
simulations in this paper. The photons are isotropically emitted
from the center $O$ (see Figure \ref{fig1}), which is the location
of the accretion disk/corona. The realistic X-ray spectrum of an AGN
is usually  more complex than a single power law and more components
will induce curvature in the spectra. However, we intend to show the
curvature produced by the torus itself and hence adopt this simple
incident spectrum. It is convenient to include more complicated
incident spectra in our simulation when we fit the observed
spectrum.

\section{Reflection spectra}

\begin{figure*}[Ht]
\includegraphics[width=0.49\linewidth]{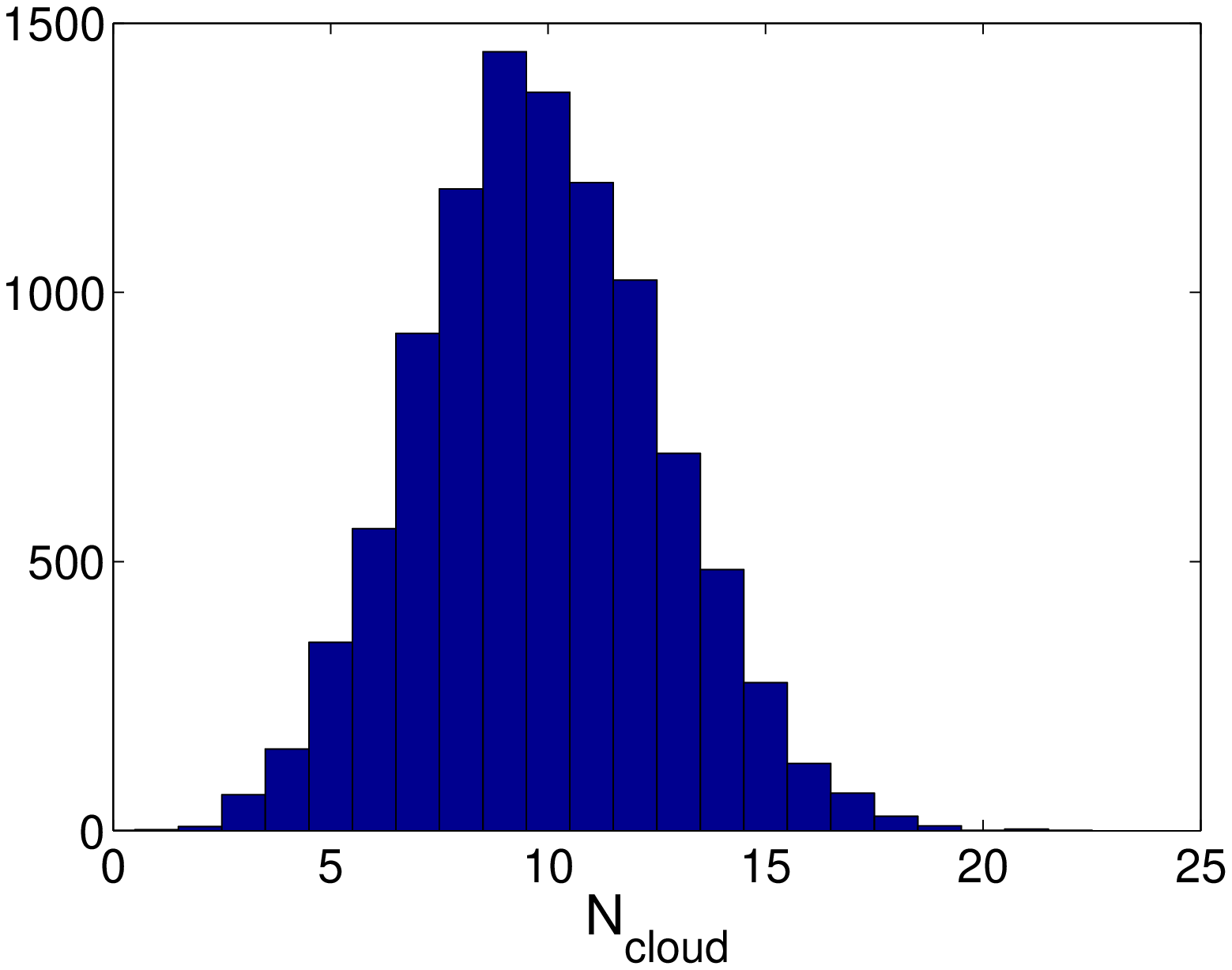}
\includegraphics[width=0.49\linewidth]{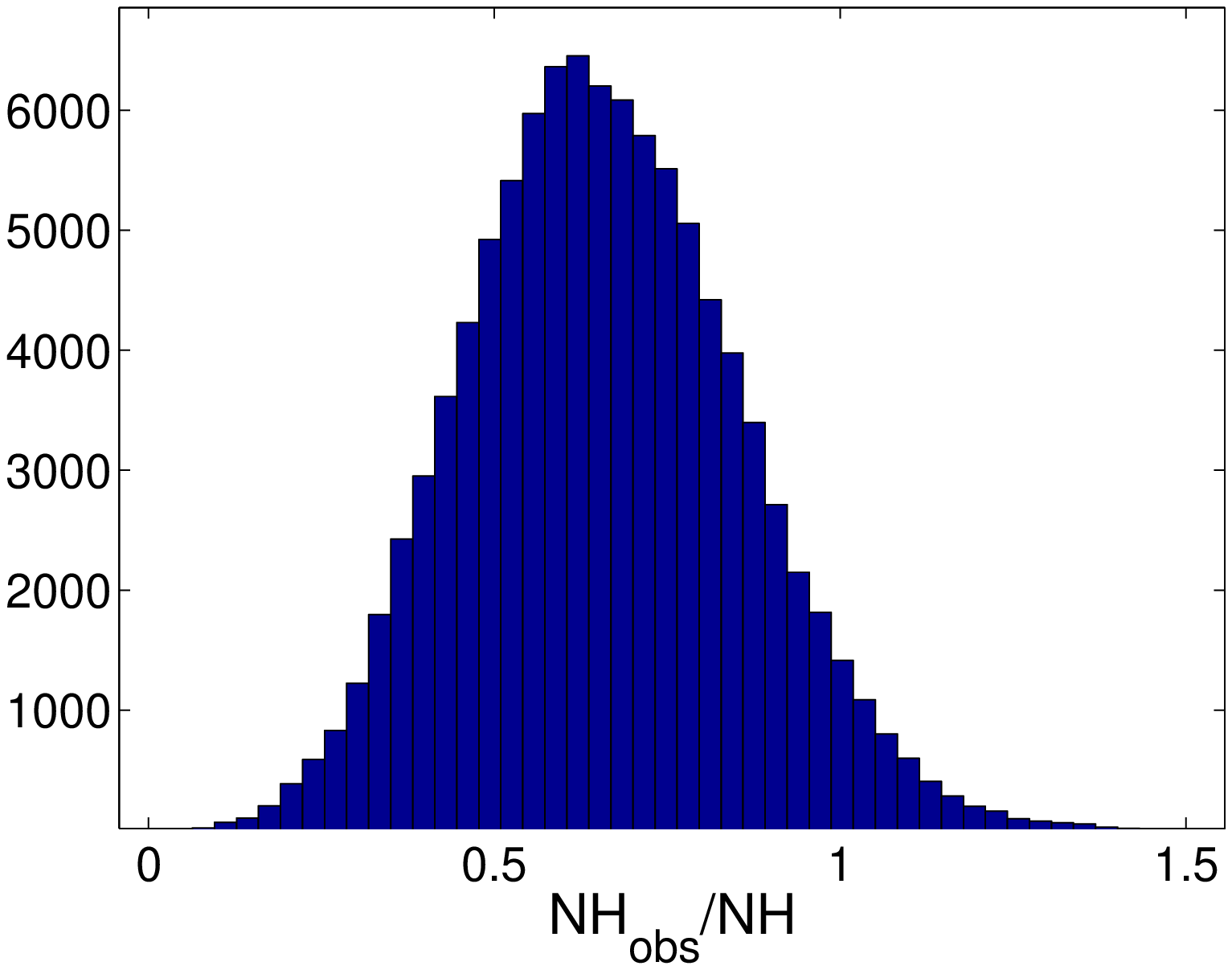}

 \caption{Statistics of the clouds for the case with $\NH=10^{24}$ cm$^{-2}$, $N=10$, and $\phi=0.01$.
 Left panel: the distribution of the number of clouds in different directions. Right
 panel:
 the distribution of the cumulative column density (normalized by $\NH$) in different directions. \label{disn}}
\end{figure*}

The $\NH$ of the clumpy torus is actually the total column density
if $N$ clouds are exactly aligned along our line of sight. However,
for a set of clouds, the number of clouds in different directions is
nearly a Poisson distribution (Nenkova et al. 2008) with a mean of
$N$ and the portion of one particular cloud along the line of sight
can be smaller than its diameter. As a result, the average column
density is smaller than $\NH$. Figure \ref{disn} (left) shows the
distribution of the number of clouds in some randomly selected
directions for $\NH=10^{24}$ cm$^{-2}$, $N=10$, and $\phi=0.01$ (the
total number of clouds in the torus is about $1.4\times10^7$ under
these parameters). The distribution of the cumulative column density
(the sum of the column density of the intersected clouds in a
particular direction) in different directions is shown in Figure
\ref{disn} (right). The mean column density (or the equivalent
$\NH$) is smaller than $\NH$ by a factor of 0.66. If we reshape the
spherical cloud into a cylinder with the same number density and
with the axis pointing to the center, the height of the cylinder
will be two-thirds of the diameter of the spherical cloud. This
``geometry-average factor'' is very close to the mean value in
Figure \ref{disn} (right). Therefore, we compare the results of the
clumpy torus with those of the smooth torus with 0.66$\NH$. We
should stress that this is not a unique method to calculate the
``equivalent $\NH$'' and there is actually no exact equivalence
between a smooth torus and a clumpy torus. We intend to present a
more meaningful comparison, but the clumpy torus is intrinsically
different from the smooth case, which is actually the motivation of
this paper.

For the smooth torus, the direct component (photons that escape from
the torus without any interaction) of the transmitted spectrum is
simply the incident spectrum weakened by the optical depth
(determined by a single $\NH$) due to photoelectric absorption and
scattering along the line of sight. For the clumpy torus, if the
size of the cloud is
 much larger than the compact corona of AGNs ($\lesssim$10 Schwarzschild radii), a single $\NH$ for the direct
 component is still appropriate, which depends on the  position of the cloud relative to the X-ray
 source (the corona). However, if our model is applied to a more extended X-ray
 source (i.e., the partial covering case), a geometry-averaged $\NH$ should be applied to the direct component,
 which depends on the density distribution within the cloud and the brightness profile of the X-ray source.
  We will explore various possibilities in the comparison with the observed spectrum in future works. Then we will only
  discuss the reflection component, i.e., the
scattered component, and the strength of the fluorescence line is
presented in the next section.

\begin{figure*}[htbp]
\plottwo{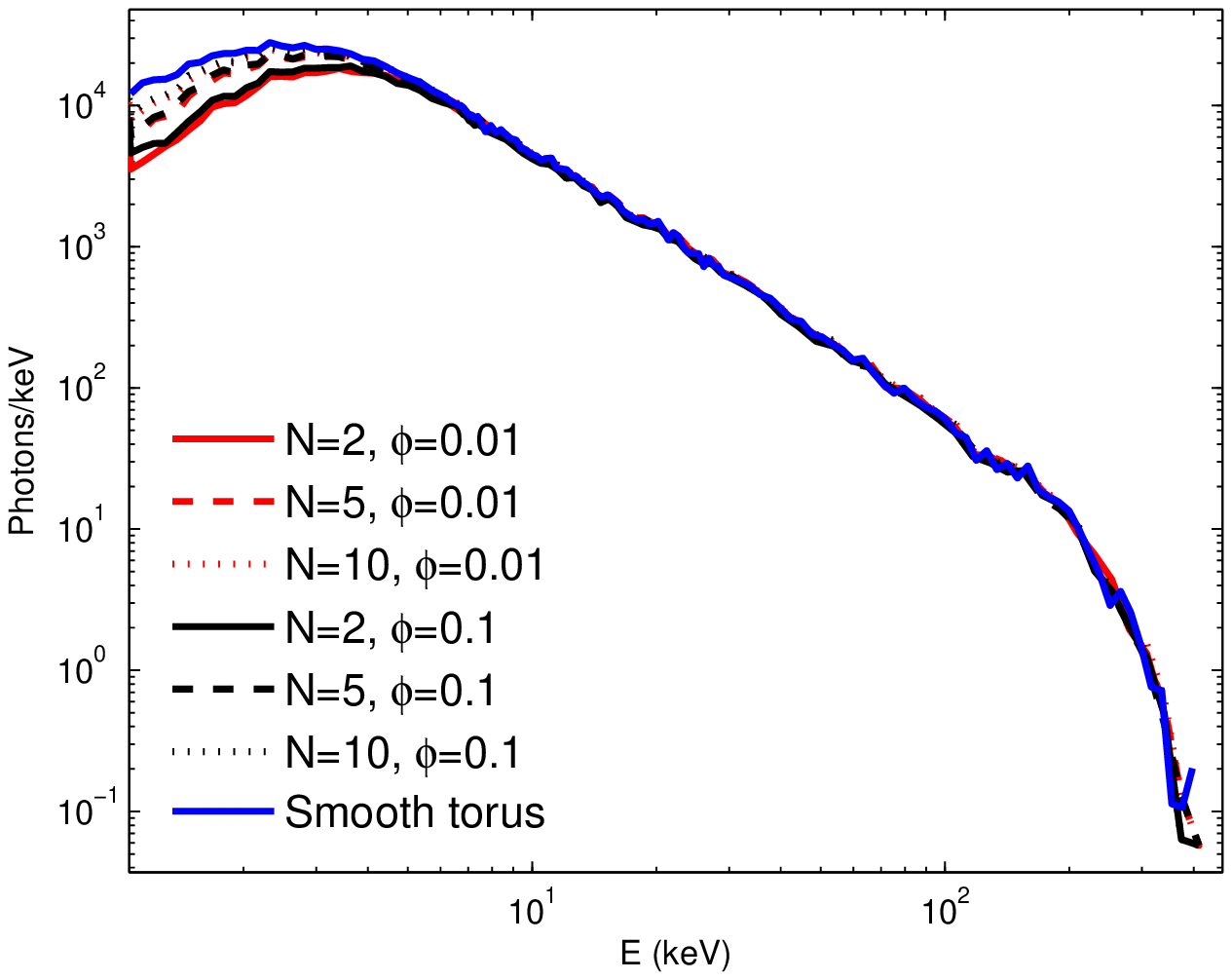}{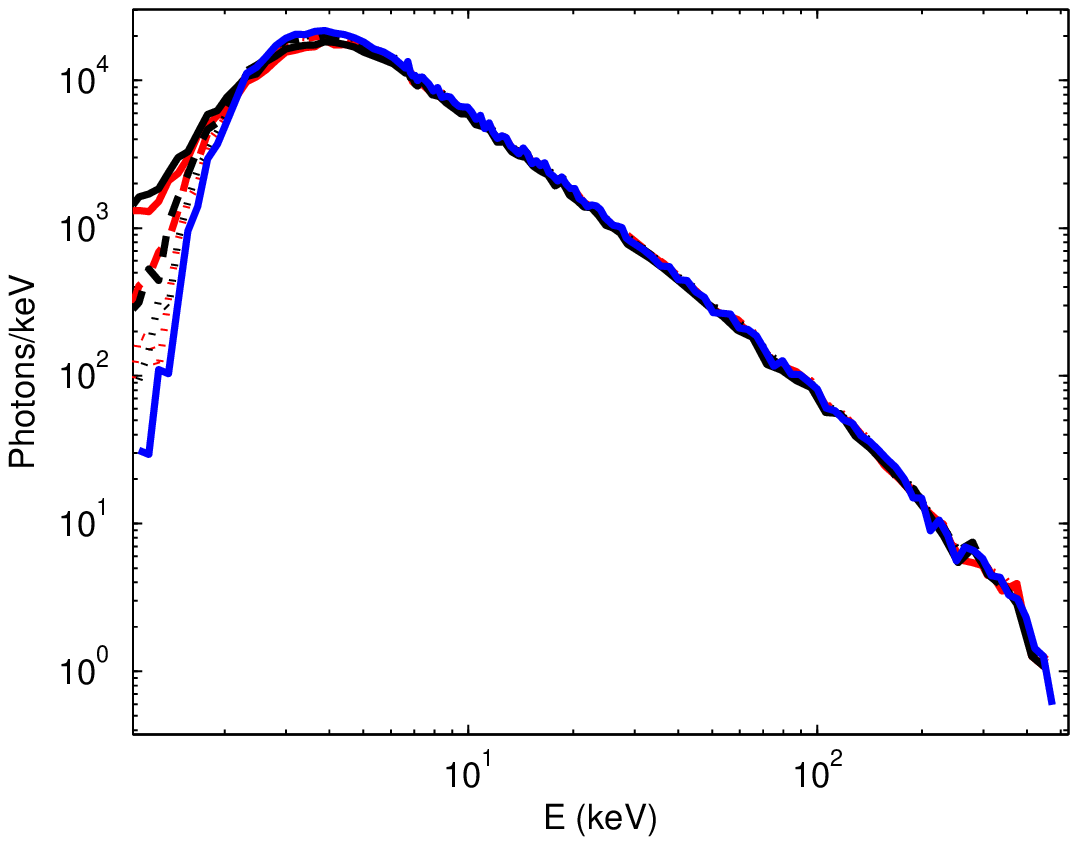}
\plottwo{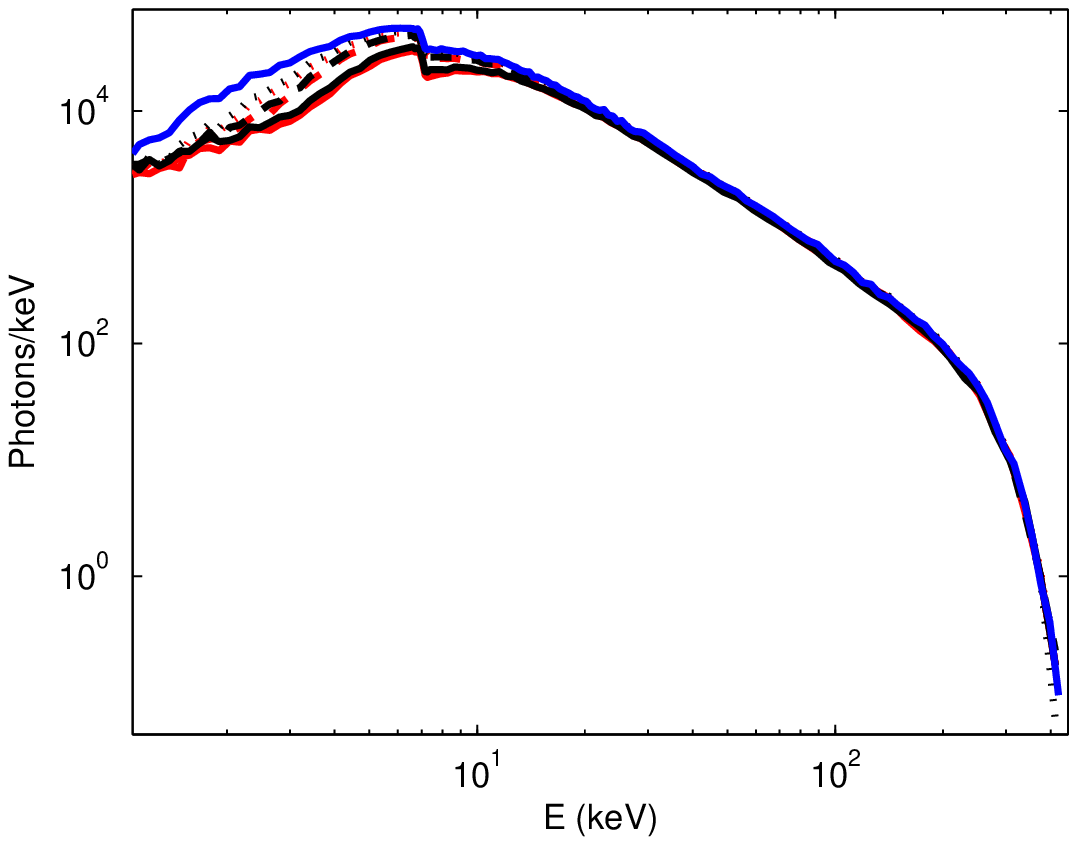}{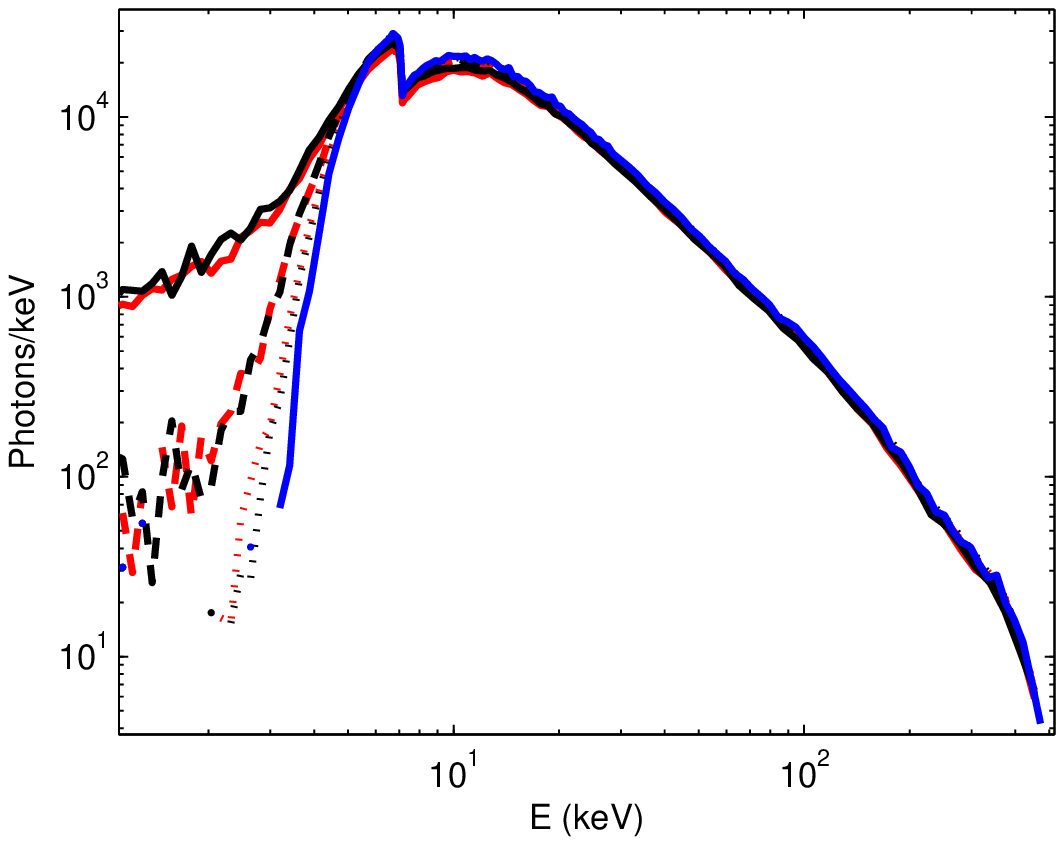}
\plottwo{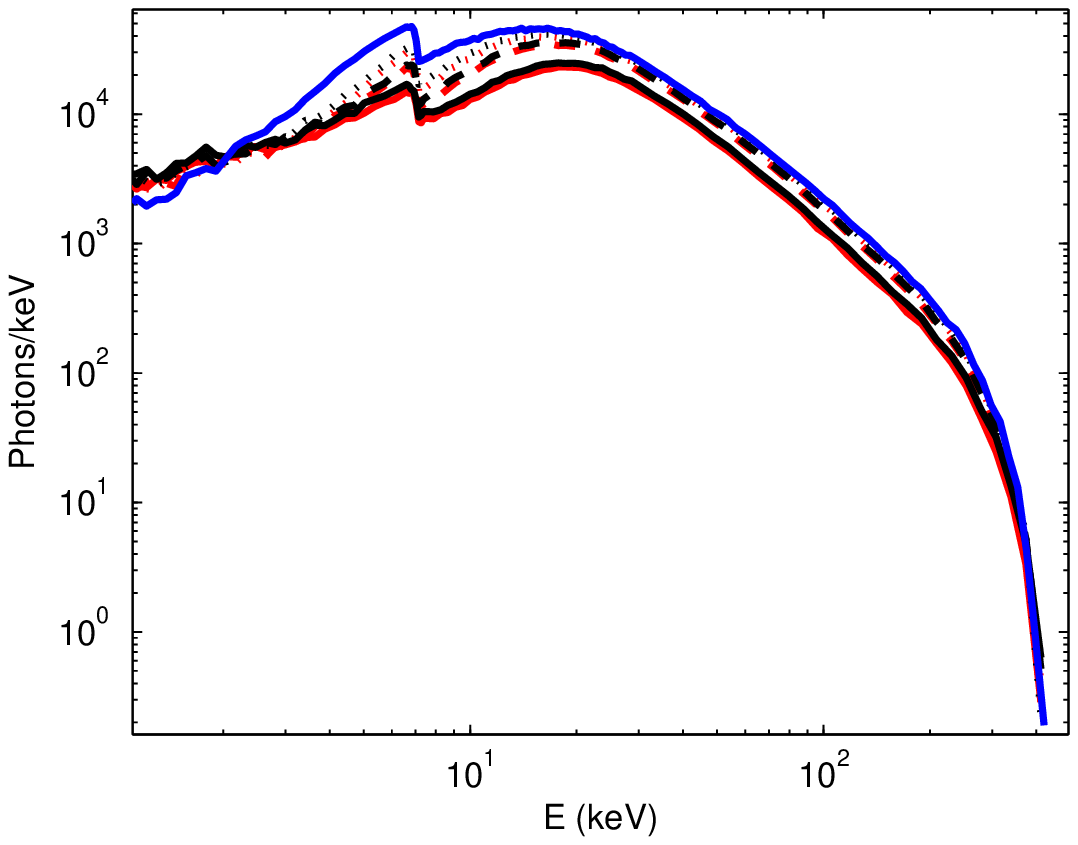}{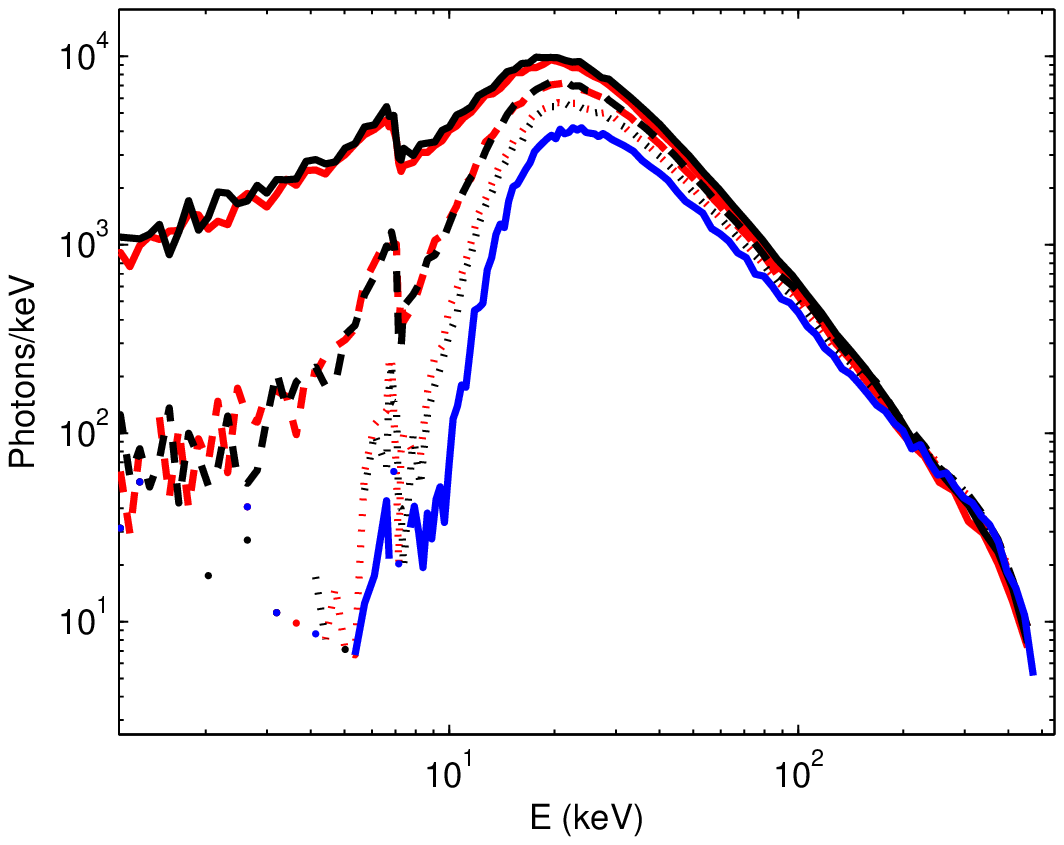}
 \caption{Reflection spectra. The top, middle, and bottom rows are for
$\NH=10^{23}$ cm$^{-2}$, $10^{24}$ cm$^{-2}$, and $10^{25}$
cm$^{-2}$, respectively. Left column: $\cos\theta=0.9-1.0$. Right
column: $\cos\theta=0-0.1$. In each panel, we show the reflection
spectra of the smooth torus with the corresponding parameters for
comparison. The meanings of curves are indicated in the legend of
the first panel and the same for each panel. \label{refl}}
\end{figure*}

\begin{figure*}[htb]

\includegraphics[width=0.33\linewidth]{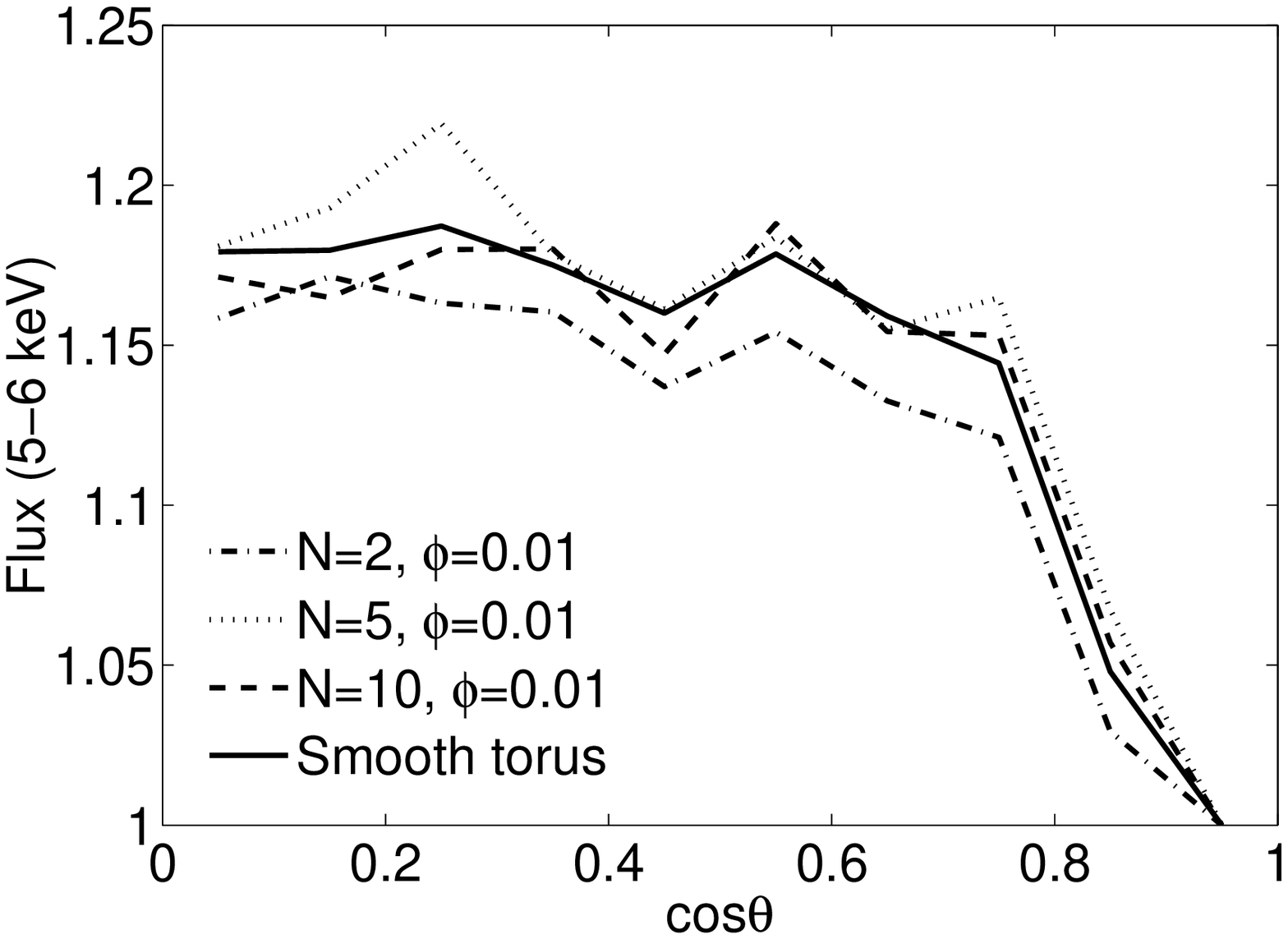}
\includegraphics[width=0.33\linewidth]{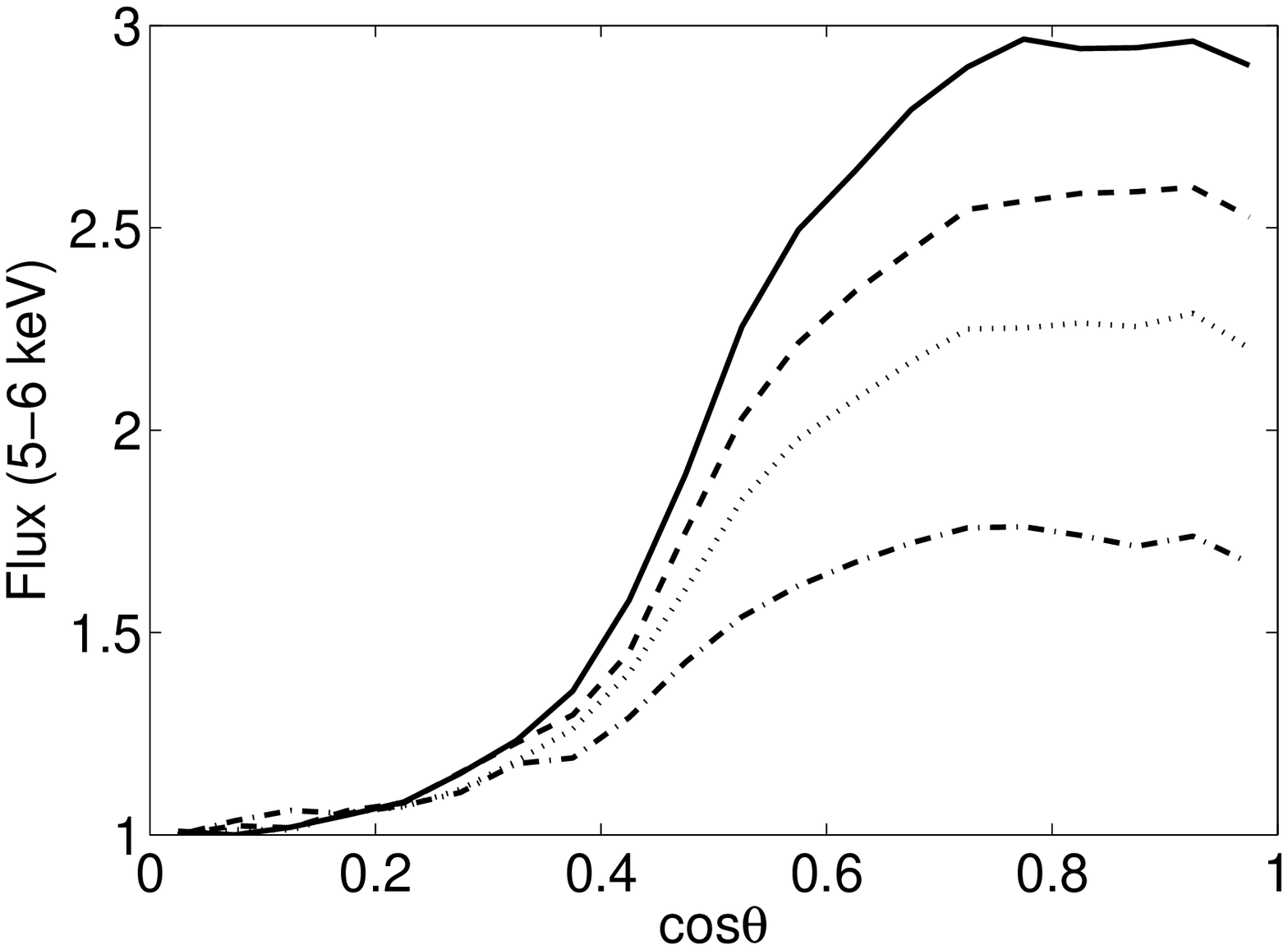}
\includegraphics[width=0.33\linewidth]{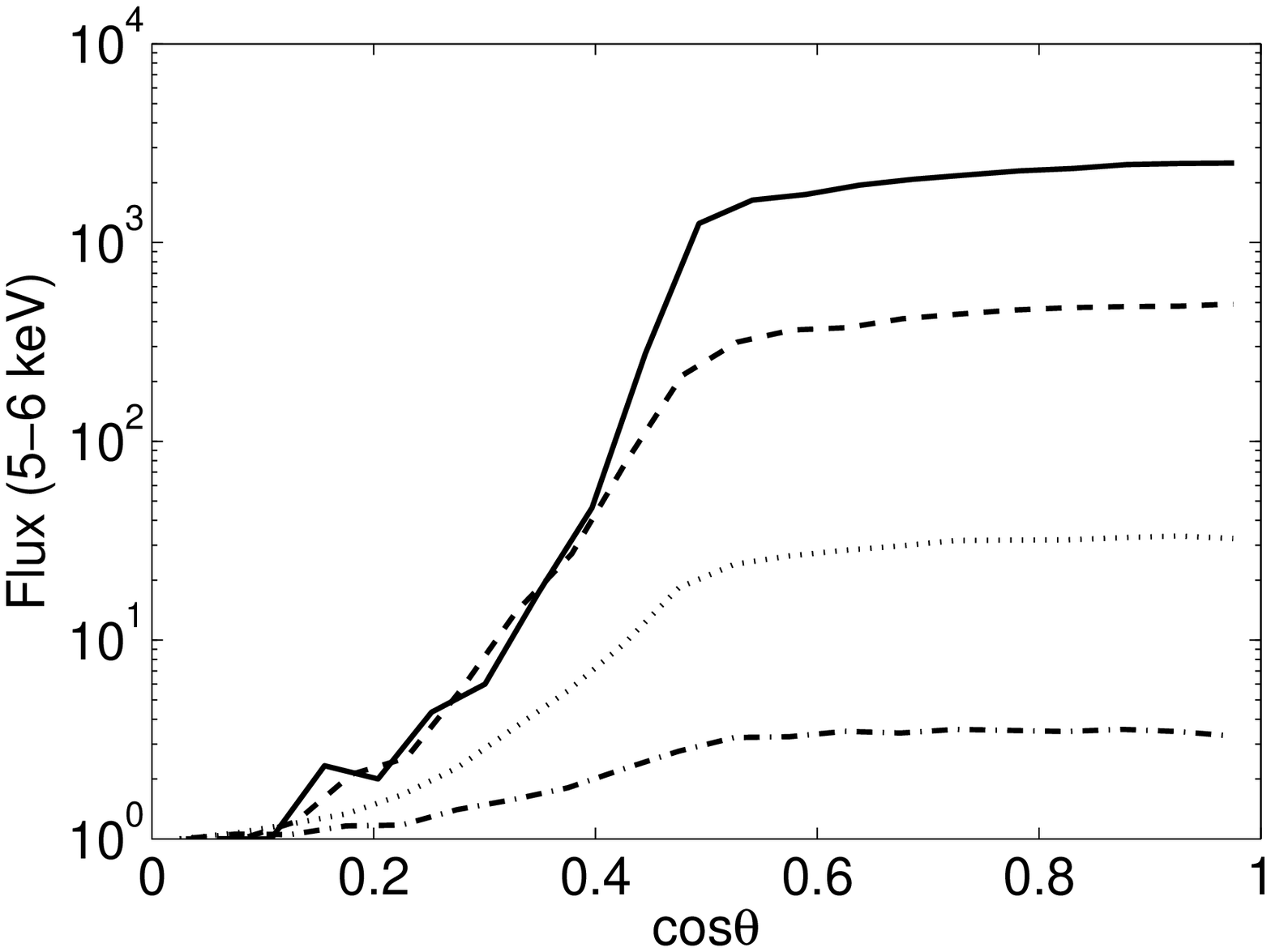}

 \caption{Distribution of the photons  in reflection spectra (5-6 keV) as a function of $\cost$ for
 $\NH=10^{23}$ cm$^{-2}$ (left), $\NH=10^{24}$ cm$^{-2}$ (middle), and $\NH=10^{25}$ cm$^{-2}$
 (right, logarithmic scale). The meanings of curves
are indicated in the legend of the first panel and the same for each
panel.
  \label{dis_con}}
\end{figure*}

In the following discussion, we  divide the direction of the photons
in the reflection spectra into 10 uniform bins according to $\cost$
($\theta$ is the angle between the direction of the photon and the
$z$-axis in Figure \ref{fig1}). The bin $\cost=0-0.1$ and
$\cost=0.9-1.0$ are defined as the edge-on and face-on directions,
respectively.
 In Figure \ref{refl}, we show the reflection spectra for different $\NH$, $N$, $\phi$,
 and $\cost$. For clarity, only the spectra of edge-on and face-on
 cases are shown.  The results of the smooth torus are also
 plotted for comparison. Next, we discuss the effect of the three clumpiness parameters
  ($\NH$, $\phi$, and $N$).
The general effect of $\NH$ is similar to the smooth case, e.g.,
 the Compton hump and the anisotropy of the reflection spectra become more
  evident with increasing $\NH$.
  The curvature above 200 keV is due to the decrease of the cross section
  of Compton scattering.
Since the column density of one cloud is solely determined by $N$
and $\NH$, the filling factor $\phi$ only slightly impacts the
spectra. However, the number of clouds along the line of sight
significantly changes the column density of one cloud and further
the shape of the spectra. As more low-energy photons can escape from
the torus with a smaller $N$ (the photons scattered by the clouds in
the far side of the torus can leak from ``holes" in the near side of
the torus), the reflection spectra become more isotropic.  We show
the distribution of photons in the 5-6 keV band as a function of
$\cos\theta$ in Figure \ref{dis_con}, where the curves are
normalized at the minima. With increasing $\NH$, more photons in the
edge-on direction are absorbed in the smooth case; however, the
anisotropy of the reflection spectra is significantly weakened in
the case of $\NH=10^{25}$ cm$^{-2}$ and $N=2$.

\begin{figure*}[htp]
\plottwo{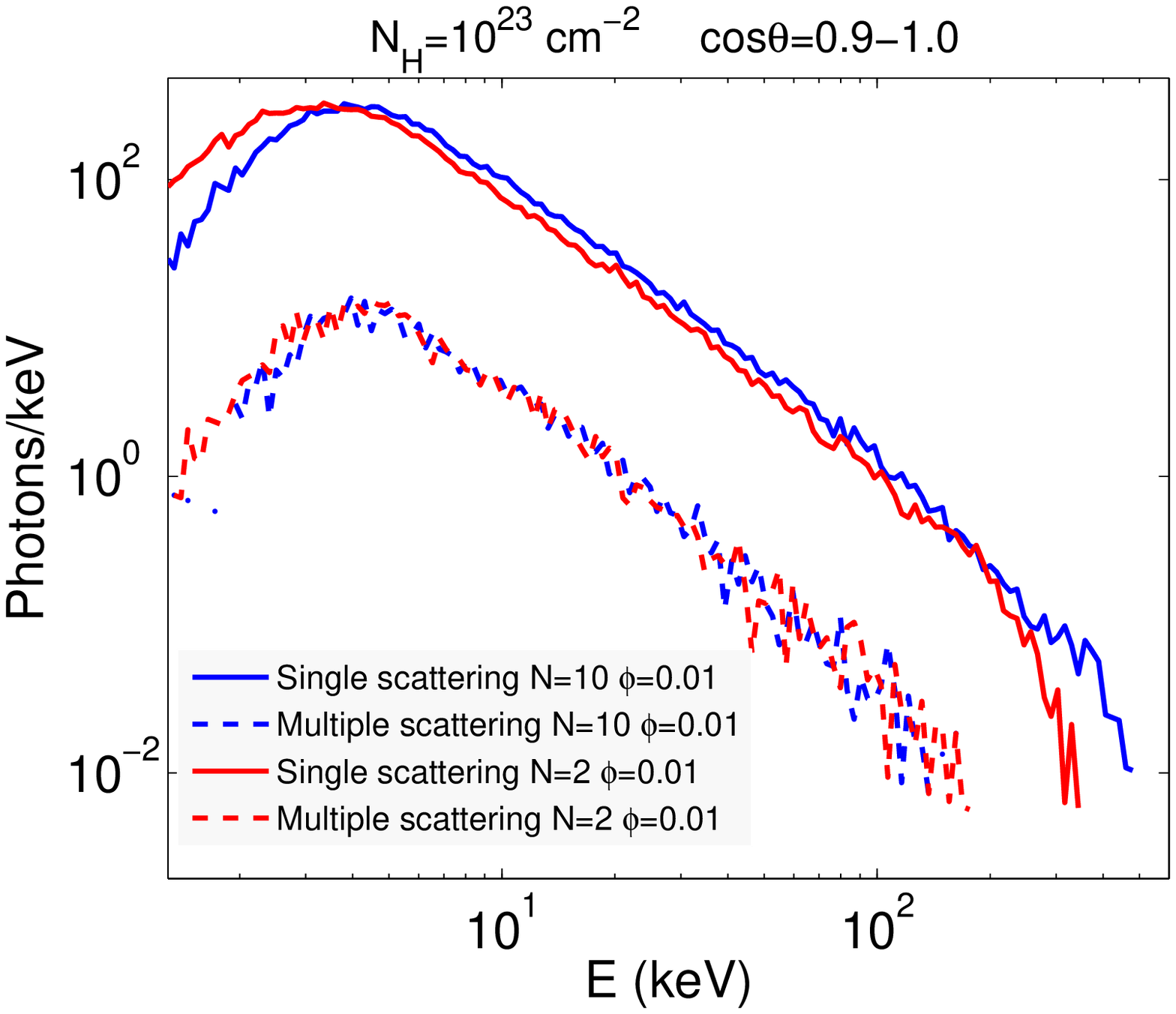}{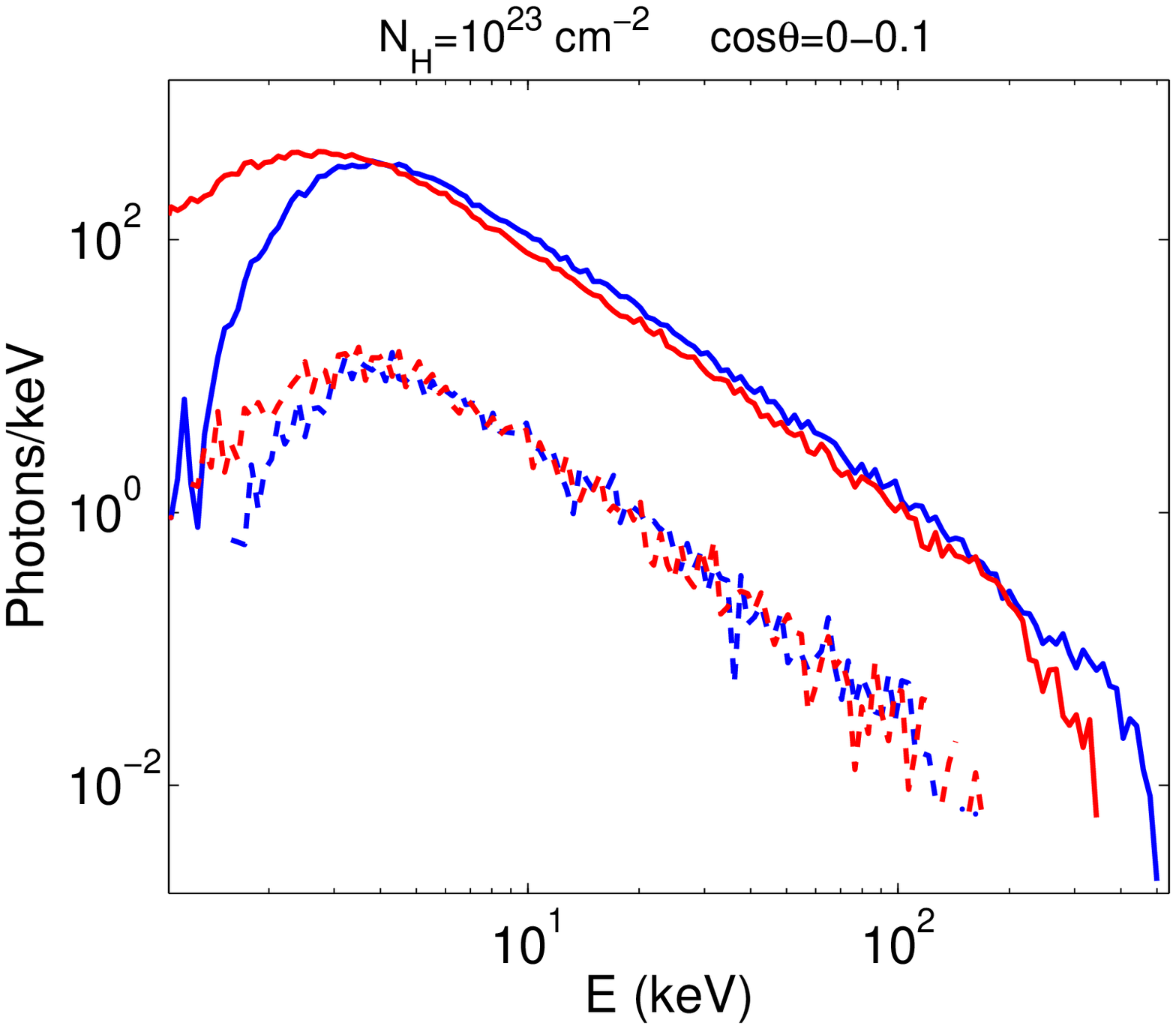}
\plottwo{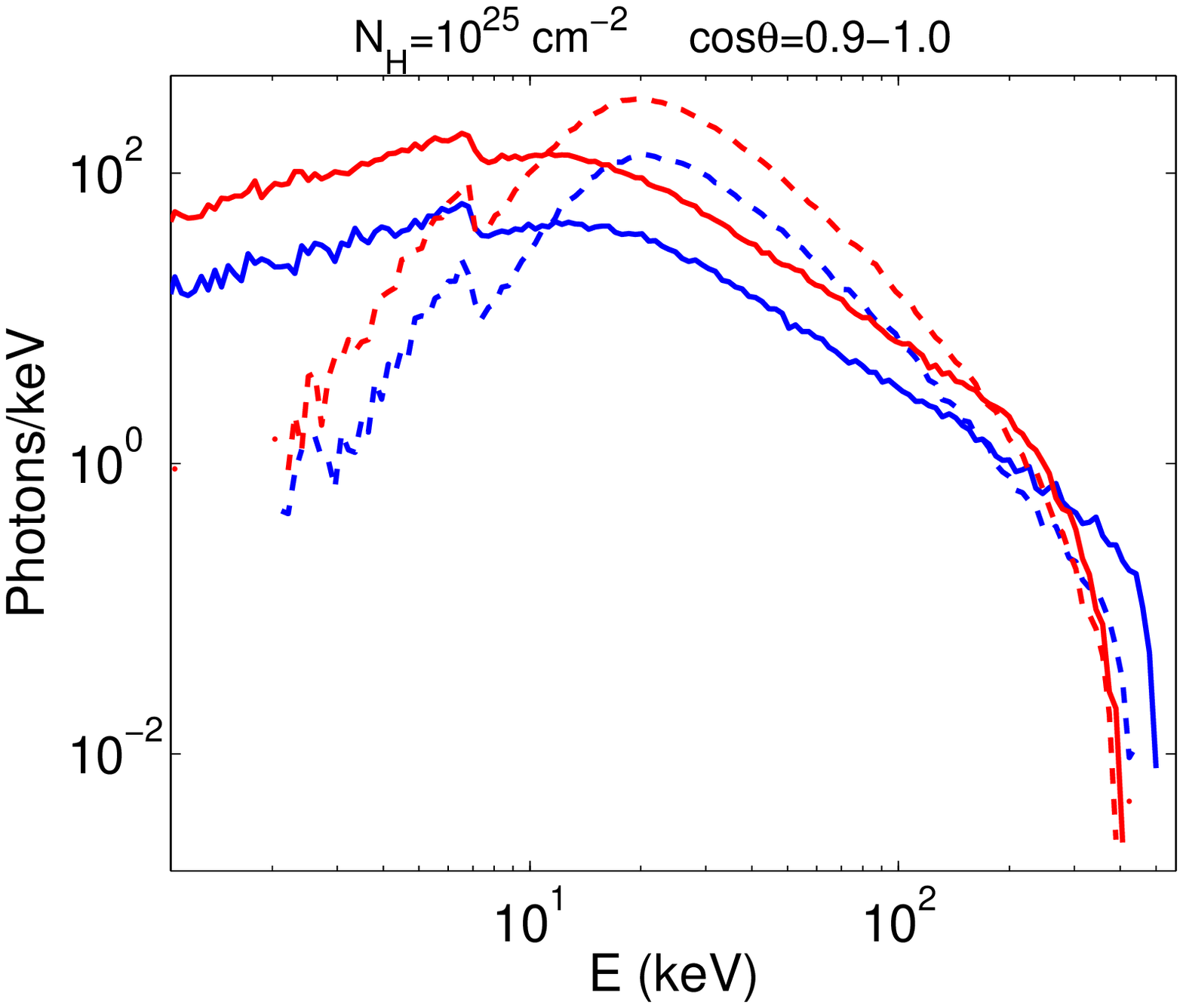}{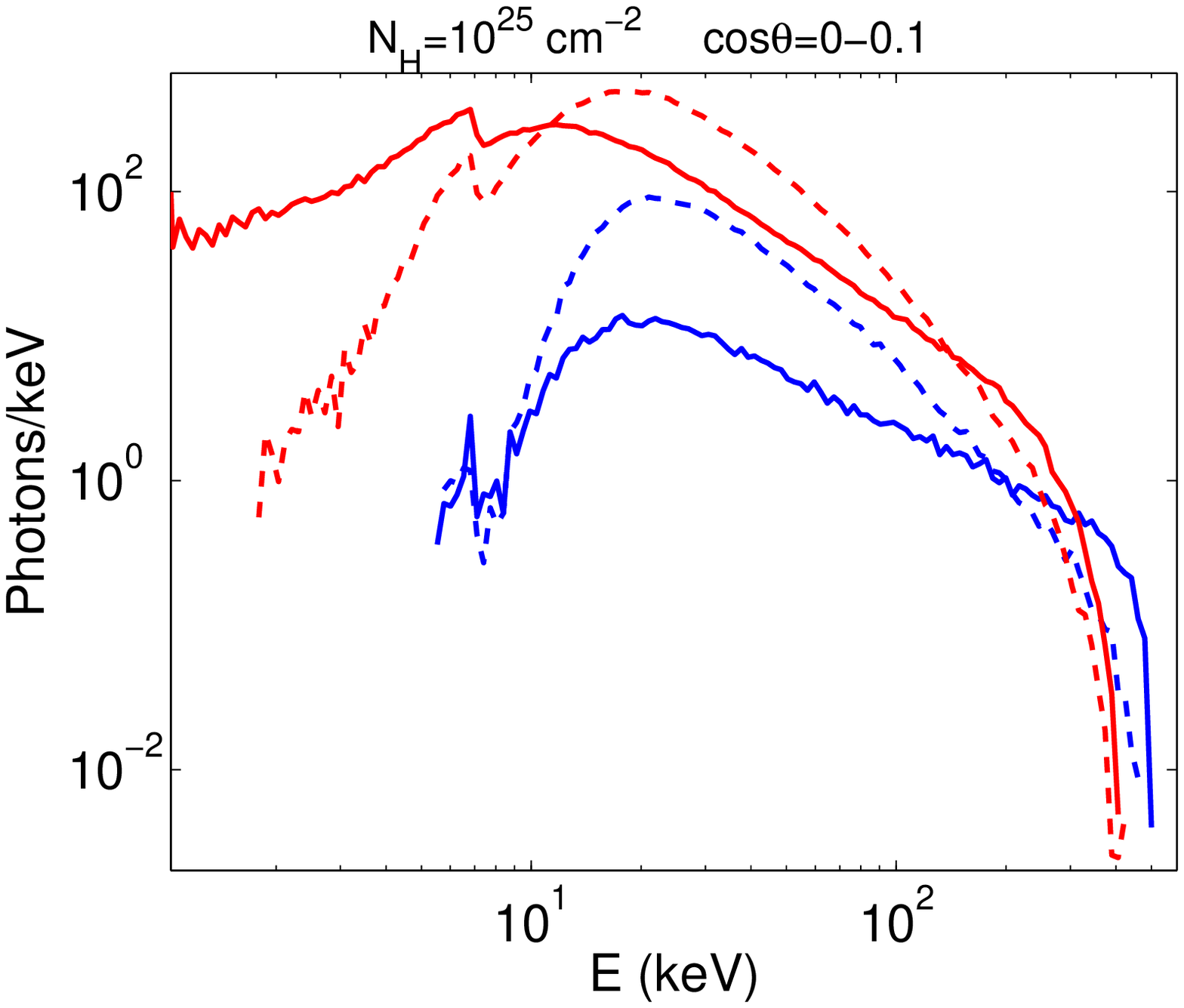} \vspace{3em}
 \caption{Single and multiple scattering spectra. The values of column density and $\cost$
are shown in the title of each panel. The meanings of the curves are
the same for each panel and
 labeled in the  first panel. \label{sing}}
\end{figure*}

\begin{figure*}[htb]
\plottwo{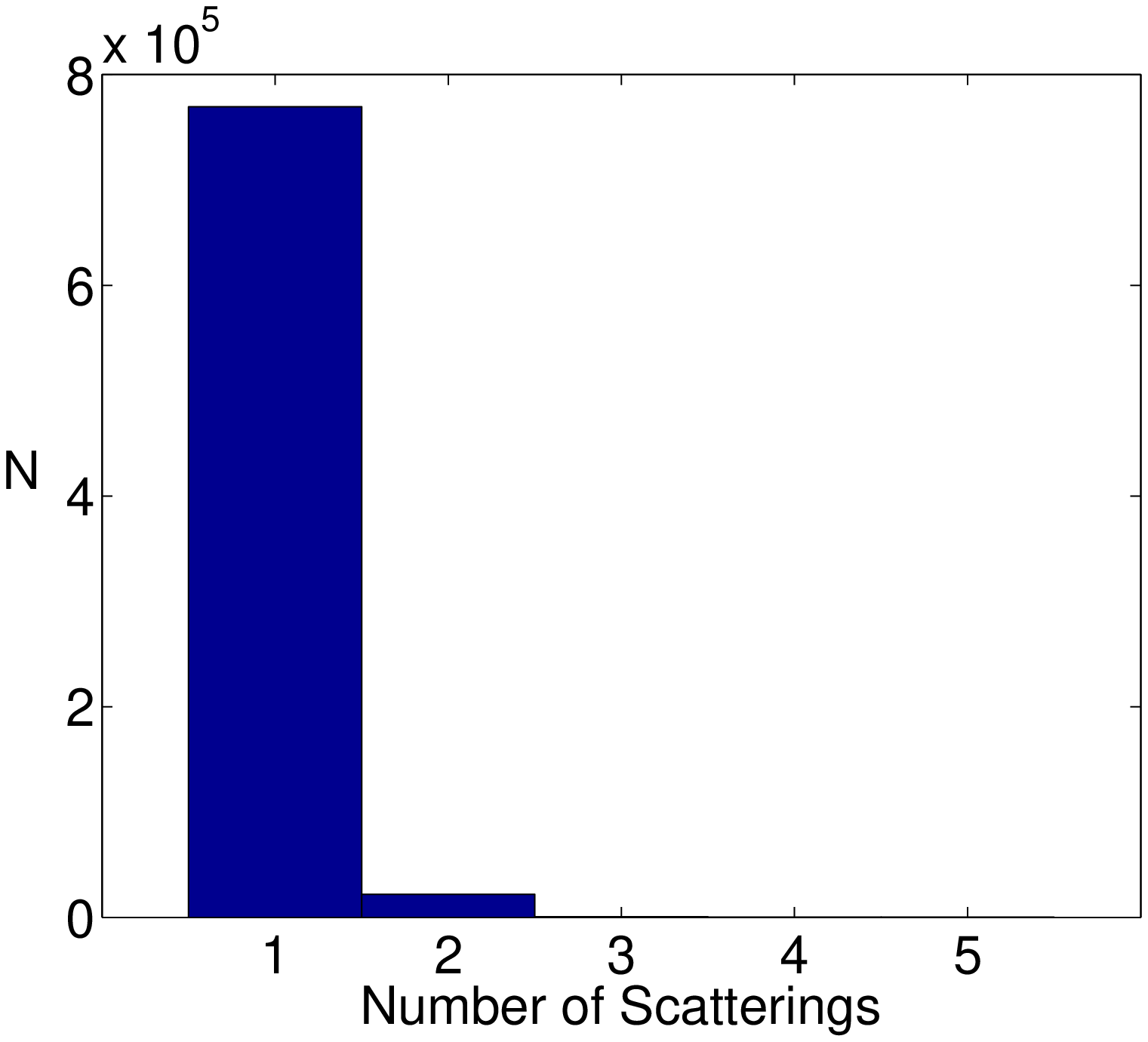}{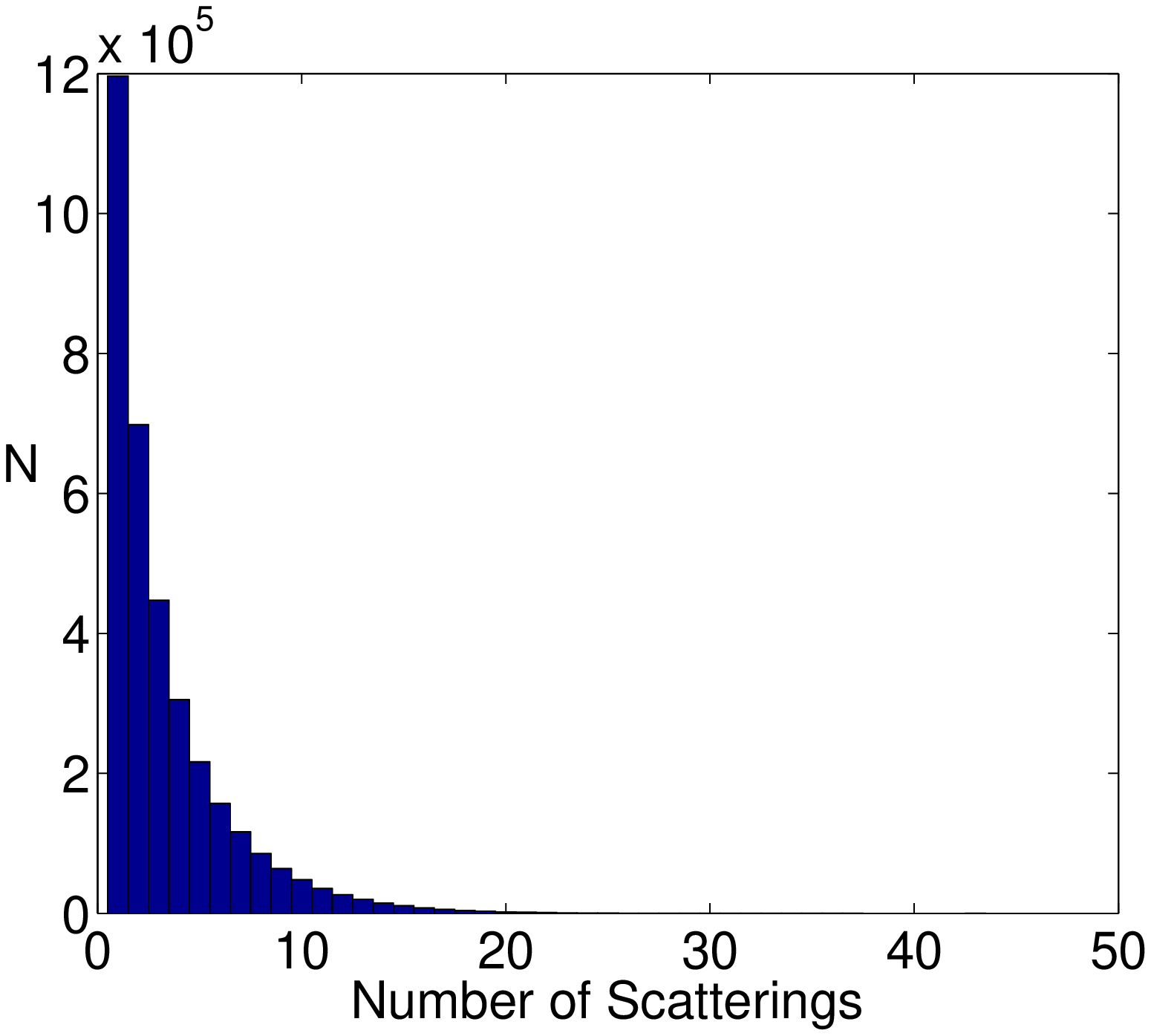}

 \caption{Histograms of the number of scatterings for $\NH=10^{23}$ cm$^{-2}$, $N=10$, and
 $\phi=0.01$ (left) and $\NH=10^{25}$ cm$^{-2}$, $N=10$, and
 $\phi=0.01$ (right). \label{sinn}
  }

\end{figure*}

To better understand the effect of $\NH$ and $N$ on the reflection
spectra, we show the single and multi-scattering spectra separately
in Figure \ref{sing}. A larger $\NH$ is helpful to suppress the
fraction of single scattering, i.e., the photons have a higher
probability of scattering with the clouds before escaping from the
torus. However, the spectrum of single scattering still dominates at
the lower energies; the multi-scattering spectrum is more important
at higher energies since it will experience more absorption during
the scatterings. The multi-scattering spectrum is more evident for
larger $N$, as there are more interfaces to produce scatterings. We
show the histogram of the number of scatterings in Figure \ref{sinn}
for $\NH=10^{23}$ cm$^{-2}$ and $10^{25}$ cm$^{-2}$. Both the
maximum number of scatterings and the fraction of multi-scatterings
increase with increasing $\NH$. In addition, since the geometry
covering factor of the clumpy torus is  $(1 - {e^{ - N}})\cos \sigma
$ (Nenkova et al. 2008), the covering factor of $N=2$ is smaller
than that of $N=10$ by a factor of 0.86.

\section{The strength of Fe K$\alpha$ line}
\begin{figure*}[htbp]
\includegraphics[width=0.33\linewidth]{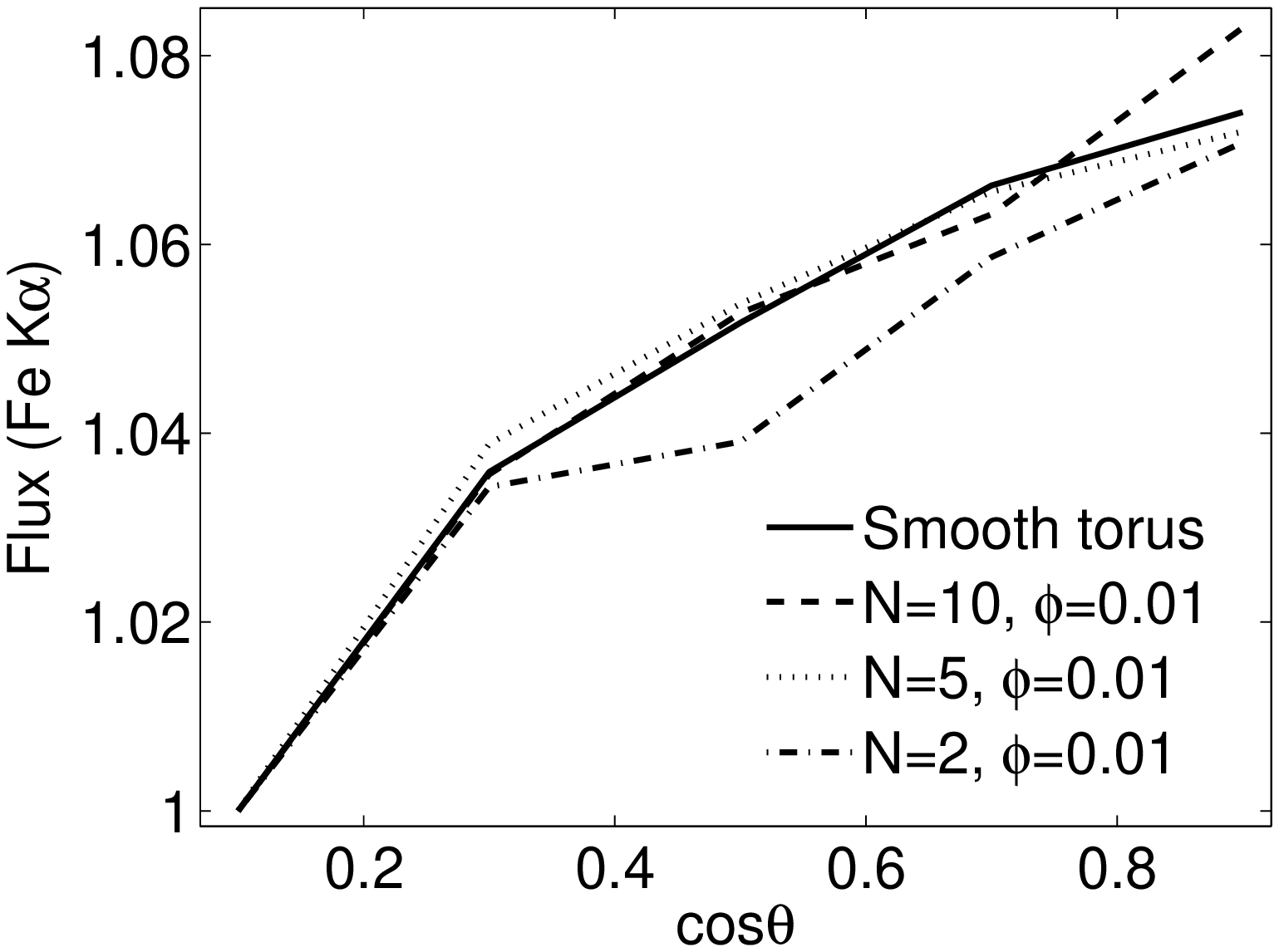}
\includegraphics[width=0.33\linewidth]{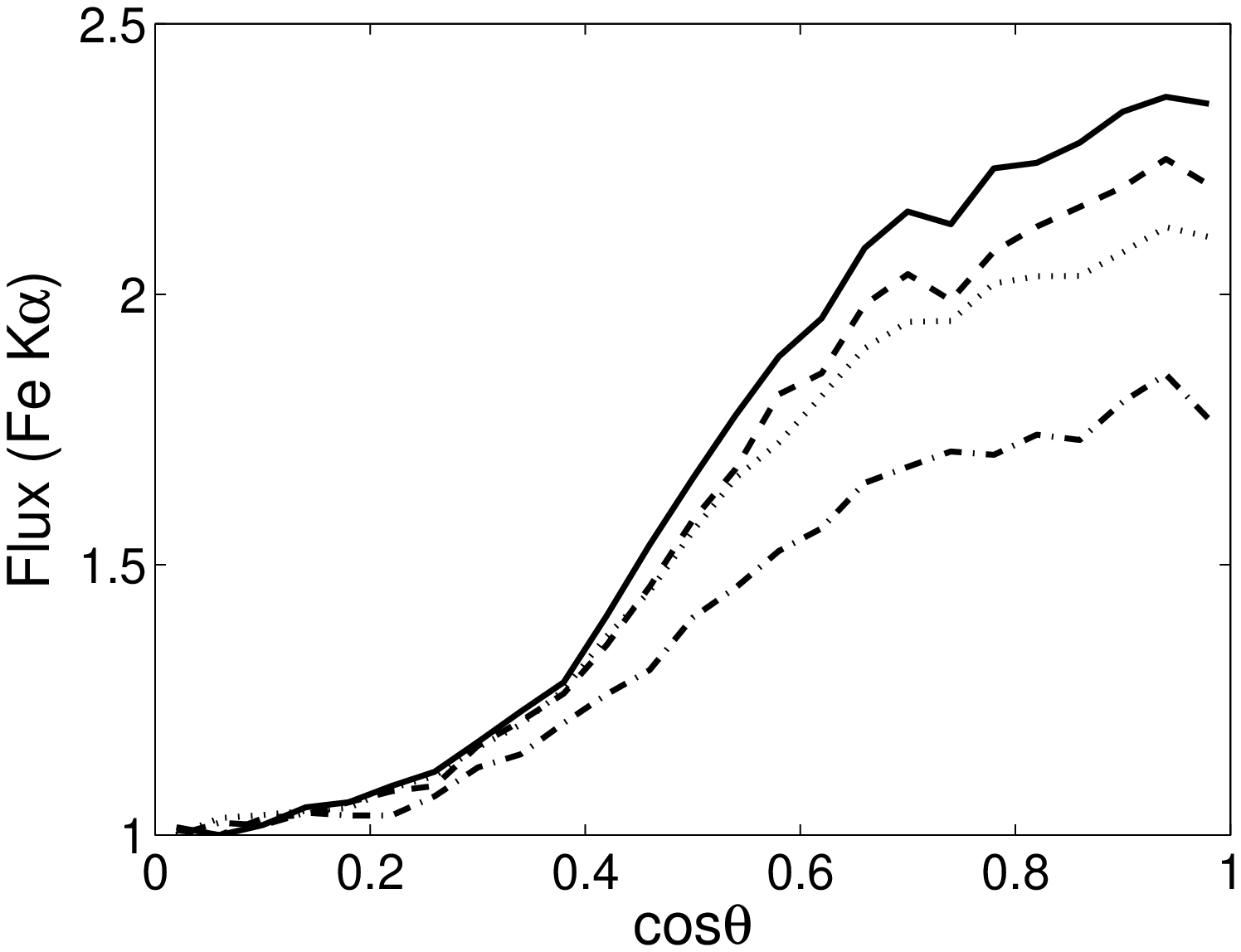}
\includegraphics[width=0.33\linewidth]{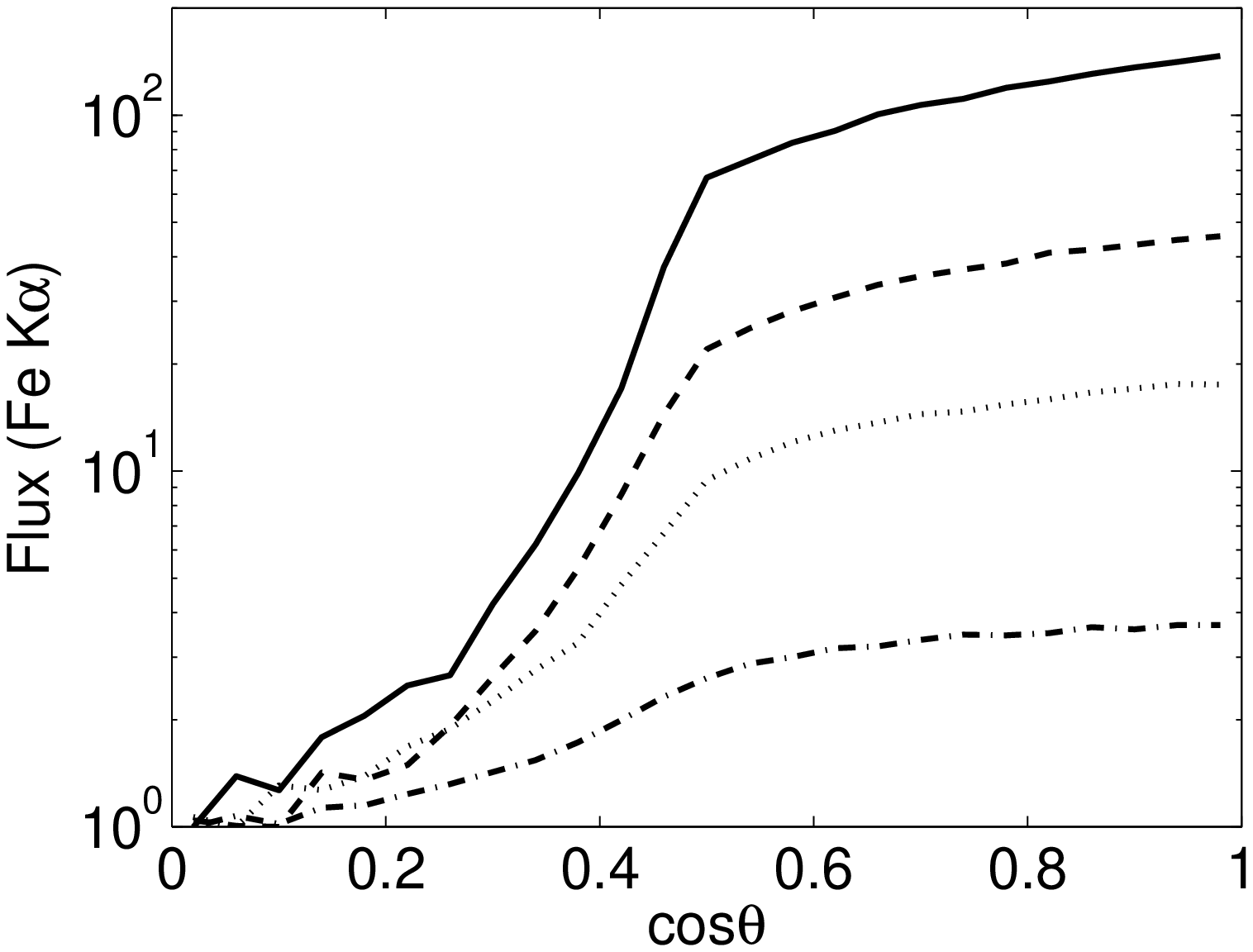}

 \caption{Distribution of the Fe K$\alpha$ photons as a function of $\cost$  for
 $\NH=10^{23}$ cm$^{-2}$ (left), $\NH=10^{24}$ cm$^{-2}$ (middle),
 and $\NH=10^{25}$ cm$^{-2}$ (right, logarithmic scale). The meanings of the curves are
 labeled in the  first panel and
the same for each panel.\label{fedis}}
\end{figure*}

The fluorescent line Fe K$\alpha$ is one of the most important lines
in the X-ray band, which can reflect the structure and density of
the torus. If the torus is optically thin to the photons of Fe
K$\alpha$, the strength of Fe K$\alpha$ can be simply calculated by
a linear relation of the properties of the torus, e.g., column
density, covering factor, and the abundance of iron (Krolik \&
Kallman 1987). Moreover, the Fe K$\alpha$ photons are isotropically
distributed. However, the optically thin approximation is not valid
for Fe K$\alpha$ photons if the column density of the torus is
larger than $2\times10^{22}$ cm$^{-2}$ (Yaqoob et al. 2010). In this
case, the geometry will impact the distribution of Fe K$\alpha$
photons and numerical simulation is required to determine the
luminosity of Fe K$\alpha$. We will investigate the relation between
the anisotropy of Fe K$\alpha$ and the parameters of clumpiness. In
the following discussion, we have included the scattered Fe
K$\alpha$ photons (the so-called ``Compton shoulder'') into the
total flux of Fe K$\alpha$. In Figure \ref{fedis}, we plot the
distribution of Fe K$\alpha$ photons as a function of $\cost$, where
the curves are normalized at the minima. Since the effect of $\phi$
is minor as shown in Figure \ref{refl}, we will only discuss the
results with different $\NH$ and $N$ but fix $\phi=0.01$. The
anisotropy increases with increasing $\NH$, which is the simple
result of more absorption in the edge-on direction in the high
column density case. For the same equivalent $\NH$, the anisotropy
of Fe K$\alpha$ in clumpy cases are weaker than the smooth case.
More specifically, a smaller $N$ will further suppress the
anisotropy of Fe K$\alpha$. This effect is much more significant in
the $\NH=25$ cm$^{-2}$ case, which is similar to the situation of
the reflection spectrum. Due to the limited statistics of the
current simulations, we can only investigate the strength of Fe
K$\alpha$. However, it is possible to extend our simulation and
discuss other fluorescence lines.

\section{Summary and discussion}
To construct an X-ray spectral model for the clumpy torus in AGNs,
we have performed simulations using an object-oriented toolkit
Geant4, by which it is convenient to deal with complex geometry.
Besides the necessary physical processes, e.g., photoelectric
absorption, Compton scattering, and fluorescence lines, considered
in previous simulations of tori,  we have included corrections to
the treatment of scattering by explicitly considering Rayleigh
(coherent) and Compton (incoherent) scattering from bound, rather
than free, electrons. This correction can induce a deviation on the
X-ray spectra up to 25\% at 1 keV, therefore we cannot ignore it in
the simulation of neutral tori. There are indications from the
widths of the Fe K$\alpha$ lines (Shu et al. 2010, 2011) that the
location of the line-emitting material may be closer to the central
engine than the traditional torus in some AGNs. Hence, the observed
effects of the corrections may vary from AGN to AGN.

Different combinations of $\NH$, $\phi$, and $N$ have been
investigated. The filling factor only slightly changes the
reflection spectra, while the number of clouds along the line of
sight significantly influences the spectra. If there are more
clouds, i.e., $N=10$, the result is similar to the smooth case;
while for the extreme case ($\NH=10^{25}$ cm$^{-2}$ and $N=2$), the
shapes of the reflection spectra in different directions are quite
similar expect for the somewhat lower amplitude in the edge-on
direction, i.e., the reflection spectra become more isotropic.
Therefore, if strong reflection components are found in the observed
spectra of type 2 AGNs, the ``clumpy" scenario could be invoked to
explain the spectra. In this case, the quantitative spectral model
presented in this paper is necessary for the measurement of the
structure of the clumpy torus.

Besides the reflection continuum, the anisotropy of the Fe K$\alpha$
line will also be impacted by clumpiness. The situation is quite
similar to the reflection continuum. In the low column density case
($\NH=10^{23}$ cm$^{-2}$), the distribution of Fe K$\alpha$ photons
is nearly isotropic and only slightly changed by the parameters of
clumpiness, while for the high column density case ($\NH=10^{25}$
cm$^{-2}$), a smaller $N$ significantly degrades the anisotropy of
Fe K$\alpha$ photons. The strength of Fe K$\alpha$ has  already been
investigated in previous simulations of smooth tori and found to be
anisotropic (though the result is expressed in the equivalent width
of Fe K$\alpha$). The clumpy torus can further smooth out the
anisotropy of the Fe K$\alpha$, which depends on the column density
of the clouds. This result is similar to the explanation of the
weakness of the silicate feature in the infrared spectra of AGNs
(Nenkova et al. 2002).

In the observational aspect, since the X-ray continua of AGNs can be
contaminated by other components not related to the torus, the
luminosity of the Fe K$\alpha$ line will provide independent
evidence of the structure of the torus.  For example, if there is no
significant difference between the luminosities of Fe K$\alpha$
lines in type 1 and 2  ($\NH>10^{23}$ cm$^{-2}$) AGNs, the ``clumpy"
torus is required according to the curves in Figure \ref{fedis} or
we should modify the unified model of AGNs as claimed by Elitzur
(2012). The current observations have already provided some clues
but are not conclusive. Liu \& Wang (2010) found that the
luminosities of the narrow Fe K$\alpha$ lines in Compton-thin and
Compton-thick type 2 AGNs are weaker than those in type 1 AGNs by a
factor of 2.9 and 5.6, respectively. This difference is broadly
consistent with the results for a smooth torus. However, from a
smaller sample from \textit{Chandra HETG}, Shu et al. (2011) found
the Fe K$\alpha$ line flux of type 2 AGNs is only marginally lower
than that of type 2 AGNs, which will require the smoothing effect of
a clumpy torus since the observed column density of Compton-thick
AGN is already larger than $10^{24}$ cm$^{-2}$. A more complete
sample of the luminosity of Fe K$\alpha$ should be helpful in
determining the geometry of a torus and the curves presented in
Figure \ref{fedis} will further constrain  the parameters.

In principle, it is possible to combine the infrared spectral energy
distribution and X-ray spectra to better constrain the structure of
tori. However, it should be cautioned that the gas in a torus is
only sensitive to X-ray photons but the dust can also absorb optical
photons. Therefore, if the dust-to-gas ratio is not uniform in a
torus (which is likely to be the case due to the temperature
gradient in the torus), the structure of the torus probed by the
X-ray photons can be different from that obtained from the infrared
spectral energy distribution. We only present the X-ray spectral
model for clumpy tori here, and the details of the comparison
between the structure from X-ray and infrared bands will be
presented in future works. The results from temporal and
polarization observations should be further helpful to break the
degeneracy of the clumpiness parameters of the torus (H\"{o}nig \&
Kishimoto 2010; Ramos Almeida et al. 2011).

\acknowledgments The authors thank the referee for useful comments
which clarified the paper. This work is supported by 973 Program of
China under grant 2009CB824800, and by the National Natural Science
Foundation of China under grant Nos. 11103019, 11133002, and
11103022.

\end{CJK*}

\end{document}